\documentclass[aps,pra,showpacs,twocolumn,superscriptaddress,floatfix]{revtex4-1}

\usepackage[colorlinks,bookmarks=false,citecolor=blue,linkcolor=red,urlcolor=blue]{hyperref}
\usepackage{epsfig}
\usepackage{tikz}
\usepackage{graphicx}

\usepackage{dcolumn}
\usepackage{bm}

\usepackage{amsmath,amssymb}

\begin{document}

\title{Generating resonating valence bond states through Dicke subradiance}
\author{R. Ganesh}
\affiliation{The Institute of Mathematical Sciences, C I T Campus, Chennai 600 113, India}
\affiliation{Homi Bhabha National Institute, Training School Complex, Anushakti Nagar, Mumbai 400094, India}
\author{L. Theerthagiri}
\affiliation{The Institute of Mathematical Sciences, C I T Campus, Chennai 600 113, India}
\affiliation{Department of Theoretical Physics, University of Madras,
Guindy Campus, Chennai 600 025, India}
\author{G. Baskaran}
\affiliation{The Institute of Mathematical Sciences, C I T Campus, Chennai 600 113, India}
\affiliation{Perimeter Institute for Theoretical Physics, Waterloo, ON, N2L 2Y6 Canada}
\date{\today}

\begin{abstract}
Dicke's original thought experiment with two spins (two-level atoms) coupled to a photon mode has recently been experimentally realized. We propose extending this experiment to many spins as a way to synthesize highly entangled states. We suggest a protocol in which we start with a direct product state of $M$ atoms in the excited state and $N$ atoms in the ground state, placed within a lossy cavity. A null observation for photon emission collapses the system onto a dark state which, remarkably, has resonating valence bond (RVB) character. 
We demonstrate this by taking advantage of the symmetry of the initial state under permutations of the $M$ excited atoms and of the $N$ unexcited atoms. Using angular momentum analysis, we reexpress the wavefunction of the dark state to illustrate its RVB character. We discuss two limiting cases in detail, with $M=1$ (one excited atom) and $M=N$ (equal number of excited and unexcited atoms). In the latter case, we show that the probability for null emission scales as $N^{-1}$ making it possible to generate highly entangled RVB states of 20 spins or more. 
\end{abstract}
\maketitle

\section{Introduction:}Dicke, in his seminal 1954 paper\citep{Dicke1954}, showed that two-level systems coupled to a common photon mode will decay coherently. 
He begins his argument with a classic thought experiment considering two two-level systems
-- one in the excited state and the other in the ground state. Na\"ive intuition suggests that this system will emit a single photon. However, by mapping the two-level systems to spin-$1/2$ spins,  the initial state is seen to be a linear combination of a dark singlet state and a bright triplet state. Hence, a photon will only be emitted with probability half. Building on this two spin experiment, Dicke argued that outgoing radiation from an N-spin system will be dominated by coherent photons emitted by `superradiant' states which have maximal total angular momentum, $S$.

The notion of superradiance has been realized and tested in many physical contexts, e.g., Bose-Einstein condensates\cite{Inouye1999}, nuclear spins\cite{Lambshift2010}, magnons\cite{Rezende2009}, excitons in quantum dots\cite{QuantumDot}, etc. Recently, Mlynek et al\cite{Mlynek2014} have recreated Dicke's original thought experiment in the lab using two superconducting qubits in a microwave cavity\cite{Blais2004}. By measuring the density matrix of the emitted photon, they showed that the spins form a mixed state with dark and bright components. Their result may be expressed as follows: two otherwise isolated spins become entangled due to their coupling to a common photon field. From this perspective, this experiment heralds a new method to create entangled quantum states in the lab.

Generating entangled states has been a long standing goal. There are well established ways to generate entangled pairs using polarized photons\cite{Ou1988}. However, it remains a challenge to generate multi-particle states with entanglement despite some recent successes\cite{Gao2010,Monz2011,Wang2016}. Several studies have examined whether Dicke superradiance could be used to generate entanglement\cite{Harkonen2009,Maniscalco2009,Wolfe2014}. 
In contrast, `subradiant' states have recently evoked interest as quantum memories, capable of storing information in long-lived states\cite{Scully2015,McGuyer2015,Mirza2016}.

In this letter, we present a protocol to synthesize certain subradiant states which are highly entangled. In particular, these states are realizations of resonating valence bond (RVB) physics, first discussed by Pauling in the context of Benzene\cite{Pauling1960}. RVB states are of great interest as precursors to high temperature superconductivity\cite{ANDERSON1973,BZA1987} and as examples of topological order\cite{Kivelson1987}. Clean experimental realizations of RVB states will go a long way in helping us understand these phenomena. However, synthesizing RVB states has proved extremely difficult so far\cite{Trebst2006}, successfully achieved only for 4 spins\cite{Nascimbene2012}.

\section{The Dicke model:}
Consider $Q$ identical qubits within a cavity resonator. The qubits are in near-resonance with one of the cavity modes so that all other modes may be neglected. The inter-qubit distance, $r$, is much smaller than the photon wavelength $(r \ll \lambda)$ so that their coupling constants do not incur a phase difference. At the same time, the distance is larger than the particle wavelength so that inter-qubit interactions may be neglected. Following Dicke's arguments, the $Q$-qubit system may be described using a wavefunction for $Q$ spin-$1/2$ moments, expressed in terms of total angular momentum states $\vert S_{tot}, m_{tot}\rangle$\cite{Tavis1968}. The spin operators for total spin are given by the sum of the individual qubit operators $\hat{S}_{tot}^\alpha = \hat{S}_1^\alpha + \ldots + \hat{S}_Q^\alpha$, with $\alpha = x,y,z$. Within the rotating wave approximation\cite{Jaynes1963}, this system is described by the Dicke Hamiltonian,
\begin{equation}
H = \lambda \hat{S}_{tot}^z + \omega_c a^\dagger a + g \{ \hat{S}_{tot}^- a^{\dagger} + \hat{S}_{tot}^+ a \},
 \end{equation} 
where $a$ ($a^\dagger$) represents photon annihilation (creation).

Dicke's thought experiment considers $Q=2$ spins, with one spin in the excited state, represented by $\vert\!\!\uparrow\rangle$, and the other in the ground state, $\vert\!\!\downarrow\rangle$. 
This state is an equal superposition of a singlet, $\vert s\rangle = \{ \vert\!\! \uparrow \downarrow \rangle - \vert\!\! \downarrow \uparrow \rangle \}/\sqrt{2}$, and a triplet wavefunction, $\vert t_{0}\rangle = \{ \vert\!\! \uparrow \downarrow \rangle + \vert\!\! \downarrow \uparrow \rangle \}/\sqrt{2}$. While the triplet component can emit a photon and decay to the $\vert t_{-1}\rangle$ state, the singlet component is dark. If an emitted photon is detected, this constitutes a measurement that will collapse the spin wavefunction onto the $\vert t_{-1}\rangle$ state. Conversely, a null observation for photon emission collapses the system onto the singlet state, leaving both spins entangled.

\section{Proposed protocol:}
Extending Dicke's thought experiment to $Q$ spins,  
a dark state can be isolated as follows: 
\begin{itemize}
\item Initialize spins in suitable direct product state 
\item Observe if any photon is emitted from cavity
\item If photon is detected, discard current run
\item If no photon is detected for several decay time periods (time scale set by lossiness of cavity), the spin wavefunction has collapsed onto a dark state
\end{itemize}
This protocol is suitable for the lossy cavity limit wherein the rate of photon loss from the cavity is much higher than the rate associated with atom-spin coupling. An emitted photon will `instantaneously' leave the cavity allowing for a precise measurement of emitted photons and their properties. The recent realization of Dicke's thought experiment\cite{Mlynek2014} was achieved in this regime. By choosing the initial direct-product state suitably, we can use this proposal to generate interesting dark states. In particular, the nature of the resulting dark state is uniquely determined by the initial direct product state. 

In what follows, we use permutation symmetries of the initial direct product state 
and properties of angular momentum addition to find an exact expression for the dark state. We then explicitly construct an RVB state and demonstrate that it is identical to the dark state.

\section{Initial State:}

We start with M qubits in the excited state and N in the ground state.
\begin{eqnarray}
 \vert \Psi_{initial} \rangle = \vert \underbrace{\uparrow \ldots \uparrow}_{M \mathrm{spins}}  \underbrace{\downarrow \ldots \downarrow}_{N \mathrm{spins}} \rangle 
 = \left| \begin{array}{c}
 \uparrow \ldots \uparrow \\
 \downarrow \ldots \ldots \downarrow
 \end{array}
 \right>.
\end{eqnarray}
For convenience, we have arranged the spins in two rows: the `up' spins in the top row and the `down' spins in the bottom row. We will assume that $M\leq N$; it can be easily seen that this is a necessary condition to have a dark state. 
Note that this initial state is invariant under in-row permutations, i.e., reordering of spins within each row. The problem is highly constrained by this symmetry which is preserved under time evolution. This is due to the nature of the Dicke Hamiltonian which is manifestly symmetric under any permutation of spins.
While the total dimension of the Hilbert space of $M + N$ qubits is $2^{M+N}$, we are only concerned with a much smaller space of dimension $(M +1) \times (N + 1)$. To see this, note that in-row permutation symmetry forces the top row to have the total angular momentum $S_{tot}=M/2$ and the bottom row to have $S_{tot}=N/2$. 
This stems from a general property of superradiant states: a state of $P$ spins that is symmetric under any permutation of the constituent spins has maximal $S_{tot}$ value, which is $P/2$\cite{klimov2009group}. Such states form a subspace of dimension $\{2(P/2)+1\} = P+1$.

\section{Collapse by photon observation:}
As the initial state has $m_{tot} = (M-N)/2$, we may write
\begin{eqnarray}
 \vert \Psi_{initial} \rangle = \sum_{S=(N-M)/2}^{(N+M)/2} a_{S} \vert S_{tot}=S, m_{tot} = \frac{M-N}{2} \rangle.\phantom{abc}
\label{eq.psiinit}
\end{eqnarray}
Each component of this state, indexed by $S_{tot}$, will emit a different a number of photons. By measuring the number of emitted photons, we perform a Stern-Gerlach-type measurement which collapses the wavefunction. In particular, observation of null emission collapses the wavefunction onto
\begin{eqnarray}
\vert \Psi_{dark}^{M,N} \rangle \sim \hat{P}_{S=(N-M)/2} \vert \Psi_{initial} \rangle.
\label{eq.dark}
\end{eqnarray}
Note that $\vert \Psi_{dark}^{M,N} \rangle$ has $m_{tot}=(M-N)/2$ and cannot reduce its $m_{tot}$ quantum number further. We assert that this state is also symmetric under in-row permutations. This is due to the projection operator itself 
being symmetric under any permutation of spins. This can be seen by construction,
\begin{equation}
\hat{P}_{S=\Sigma} = \prod_{S'\neq \Sigma} \frac{\hat{S}_{tot}^2 - S'(S'+1)}{\Sigma(\Sigma+1) - S'(S'+1) }.
\end{equation}
As it only contains $\hat{S}_{tot}$, it is symmetric under any permutation of the constituent spins.

 \section{Row-wise decomposition:}
We now reepxress the dark state in terms of row-wavefunctions. On the basis of the above arguments, we have
\begin{eqnarray}
\nonumber \vert \Psi_{dark}^{M,N} \rangle  = \sum_{\lambda=0}^{M} C_{\lambda} \vert S_{tot} = M/2, m_{tot} = M/2-\lambda \rangle_{t} \otimes \\
\vert  S_{tot} = N/2, m_{tot} = \lambda-N/2  \rangle_{b}.\phantom{a}
\label{eq.schmidt}
\end{eqnarray}
The subscripts `t' and `b' denote `top' and `bottom' rows respectively. The $S_{tot}$ quantum numbers are fixed to their maximal values due to in-row permutation symmetry. The $m_{tot}$ values have been assigned to conserve the z-component of total spin, $m_{tot}=(M-N)/2$. 

It is now apparent that the coefficients $C_{\lambda}$ are simply Clebsch Gordan coefficients, $C(j_1 j_2 J; m_1 m_2)$, with $j_1 = M/2$, $m_1 = M/2-\lambda$, $j_2 = N/2$, $m_2 = \lambda-N/2  $ and $J = (N-M)/2$. Using known expressions for the Clebsch Gordan coefficients\cite{Rose}, we obtain 
\begin{equation}
C_{\lambda} = (-1)^\lambda \left[ 
\frac{(N-M+1)M! (N-\lambda)!}{(N+1)!(M-\lambda)!}
  \right]^{1/2}.
  \label{eq.Clambda}
\end{equation}

\section{RVB construction:}
We now construct the conjectured RVB state and rewrite it 
in a form similar to Eq.~\ref{eq.schmidt}. By explicitly evaluating the expansion coefficients, we see that the RVB state is identical to the dark state above.

\begin{figure}
\includegraphics[width=3.3in]{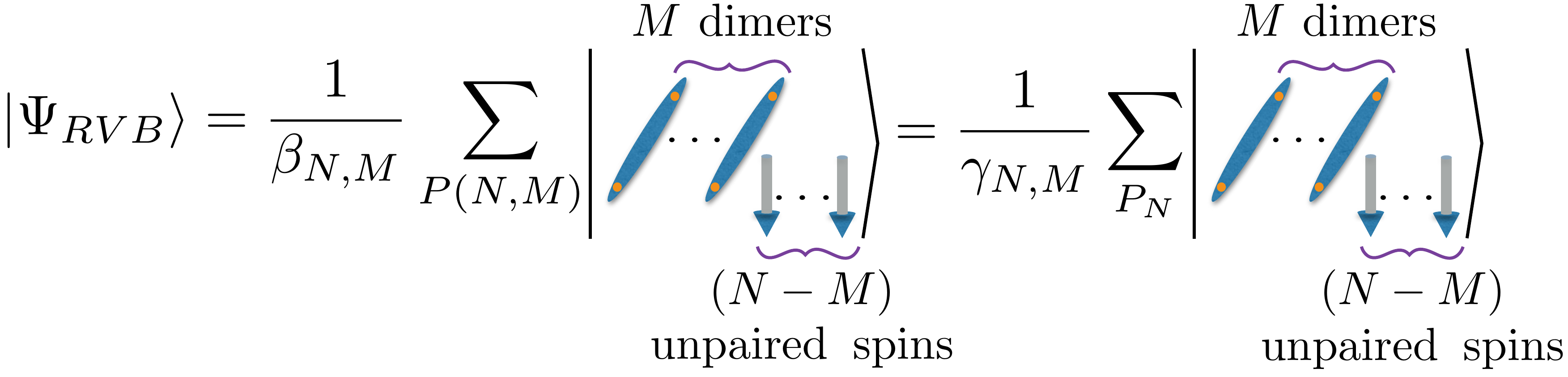}
\caption{ 
\textit{Explicit construction of the RVB state.} The sum is over $P(N,M)$ - the number of ways of assigning one partner from the bottom row for every spin in the top row, leading to ${}^N \!P_M$ terms. The quantity $\beta_{N,M}$ is a normalisation constant. Upto an overall constant, this sum can be rewritten as a sum over all permutations of spins in the bottom row, with $N!$ terms. The extra constant has been absorbed into the normalisation constant, $\gamma_{N,M}$.  
}
\label{fig.RVB}
\end{figure}

Our construction is shown in Fig.~\ref{fig.RVB}. We pair each spin from the top row with one of the spins in the bottom row. With each pair, we associate a singlet wavefunction, $\{  \vert\! \uparrow_{t} \downarrow_{b}\rangle - \vert \!\downarrow_{t} \uparrow_{b}\rangle  \}/\sqrt{2} $. We follow a fixed ordering convention similar to the Marshall sign convention in a bipartite lattice\cite{Marshall1955}: in a singlet wavefunction, the spin in the top row occurs first. 
As $M <N$, every spin in the top row is part of a singlet while $(N-M)$ spins remain unpaired in the bottom row. The pairing of spins can be done in many ways; to be precise, the number of possibilities is ${}^N\! P_M$, the number of ways of choosing an ordered set of $M$ objects out of $N$. As argued in Fig.~\ref{fig.RVB}, this can be recast as a sum over all permutations of the spins in the bottom row with $N!$ terms.  
 \textit{Our RVB state is the equal-amplitude in-phase superposition of every such pairing configuration}.

Having explicitly defined the RVB wavefunction, we expand it in the  $S_z$ basis,
to obtain 
\begin{eqnarray}
\nonumber \vert \Psi_{RVB} \rangle = \sum_{\kappa=0}^{M} \frac{(-1)^\kappa}{\gamma_{N,M} 2^{M/2}}  \left[  
\sum_{C(M,\kappa)} \vert \underbrace{\downarrow \ldots \downarrow}_{\kappa } \underbrace{\uparrow \ldots \uparrow}_{M-\kappa } \rangle_{t}
\right]
\otimes \\
\!\!\!\left[
\sum_{P_N} \vert \underbrace{\uparrow \ldots \uparrow}_{\kappa } \underbrace{\downarrow \ldots \downarrow}_{N-\kappa} \rangle_{b}
\right].\phantom{ac}
\label{eq.RVB_kappa}
\end{eqnarray}
The factor of ${2}^{-M/2}$ comes from the $\sqrt{2}$ factor in each singlet wavefunction. The summation over $C(M,\kappa)$ is over $\left(  \begin{array}{c} M \\ \kappa \end{array}\right)$ terms, corresponding to the number of ways of choosing $\kappa$ out of $M$ spins. The summation over $P_N$ denotes a sum over all permutations of the N spins in the bottom row.

The wavefunctions for the top and bottom rows above are clearly permutation symmetric. They can be written in terms of angular momentum eigenstates using 
\begin{eqnarray}
\nonumber \vert S_{tot=M/2}, m_{tot}={M/2-\kappa} \rangle_{t} =\left(  \begin{array}{c} M \\ \kappa \end{array}\right)  ^{-1/2} 
\\
\nonumber \times \sum_{C(M,\kappa)} \vert  \underbrace{\uparrow \ldots \uparrow}_{M-\kappa} 
\underbrace{\downarrow \ldots \downarrow}_{\kappa }\rangle_{t}, \\
\nonumber \vert S_{tot=N/2}, m_{tot}={\kappa-N/2} \rangle_{b} = \left(  \begin{array}{c} N \\ \kappa \end{array}\right)  ^{-1/2} \\
\nonumber \times
\sum_{C(N,\kappa)} \vert \underbrace{\uparrow \ldots \uparrow}_{\kappa } \underbrace{\downarrow \ldots \downarrow}_{N-\kappa} \rangle_{b} \\
= \frac{1}{\kappa! (N-\kappa)!} \left(  \begin{array}{c} N \\ \kappa \end{array}\right)  ^{-1/2}
\sum_{P_N}  \vert \underbrace{\uparrow \ldots \uparrow}_{\kappa} \underbrace{\downarrow \ldots \downarrow}_{N-\kappa} \rangle_{b}. \phantom{abcd}
\end{eqnarray}
The summations over $C(M,\kappa)$, $C(N,\kappa)$ and $P_N$ are as defined above.
We conclude that 
\begin{eqnarray}
\nonumber \vert \Psi_{RVB} \rangle &=& \sum_{\kappa=0}^{M} \frac{(-1)^\kappa}{\gamma_{N,M} 2^{M/2}} {\kappa! (N-\kappa)!}  
 \left(  \begin{array}{c} N \\ \kappa \end{array}\right)  ^{1/2}
   \left(  \begin{array}{c} M \\ \kappa \end{array}\right)^{1/2} \\
\nonumber &\phantom{a}&\vert S_{tot=M/2}, m_{tot}={M/2-\kappa} \rangle_{t}  \otimes \\
&\phantom{a}&\vert S_{tot=N/2}, m_{tot}={\kappa-N/2} \rangle_{b}.\phantom{acb}
\label{eq.Schmidt_RVB}
\end{eqnarray}
We have effectively performed a Schmidt decomposition into top and bottom row wavefunctions. 
We now compare this with Eq.~\ref{eq.schmidt}. We note that $\vert S_{tot}=M/2, m \rangle_{t} $ and $\vert S_{tot}=N/2, m \rangle_{b} $
are uniquely defined states -- for a system with $P$ spins, in the superradiant ($S_{tot}=P/2$) sector, states are uniquely identified by their $m_{tot}$ quantum numbers (see Appendix). 
If we choose the normalisation constant to be
\begin{equation}
\gamma_{N,M} = \frac{N!}{2^{M/2}} \sqrt{\frac{N+1}{N-M+1}},
\end{equation} 
we see that the coefficients in Eqs.~\ref{eq.schmidt} and \ref{eq.Schmidt_RVB} are identical. We have shown that the row-wise expansions of $\vert \Psi_{dark}^{M,N}\rangle$ and $\vert \Psi_{RVB}\rangle$ have identical coefficients. Since the basis states in the decomposition are unique,
it follows that 
\begin{equation}
\vert \Psi_{dark}^{M,N}\rangle = \vert \Psi_{RVB}\rangle.
\end{equation}
The subradiant state realized by a null-measurement is, in fact, the RVB state.

\section{Probability of null emission:}
The probability for null emission (i.e., for collapse onto the RVB state) is $\vert a_{(N-M)/2} \vert^2$, where $a_{(N-M)/2}$ is the amplitude of the 
dark component in Eq.~\ref{eq.psiinit}. We have $ a_{(N-M)/2} = \langle \Psi_{initial} \vert \Psi_{RVB}\rangle$. Using the form of $\vert \Psi_{RVB}\rangle$ in Eq.~\ref{eq.RVB_kappa}, we see that only $\kappa=0$ contributes to the overlap. We obtain 
\begin{eqnarray}
 \mathrm{P}(\mathrm{null\phantom{a}emission})
\!= \!\vert \langle \Psi_{initial} \vert \Psi_{RVB}\rangle \vert^2\!= \!{\frac{N-M+1}{N+1}}.\phantom{abce}
\label{eq.adark}
\end{eqnarray}
We now consider two special cases.

\section{Initial state with single excitation:}
\begin{figure}
\centering
\includegraphics[width=3.3in]{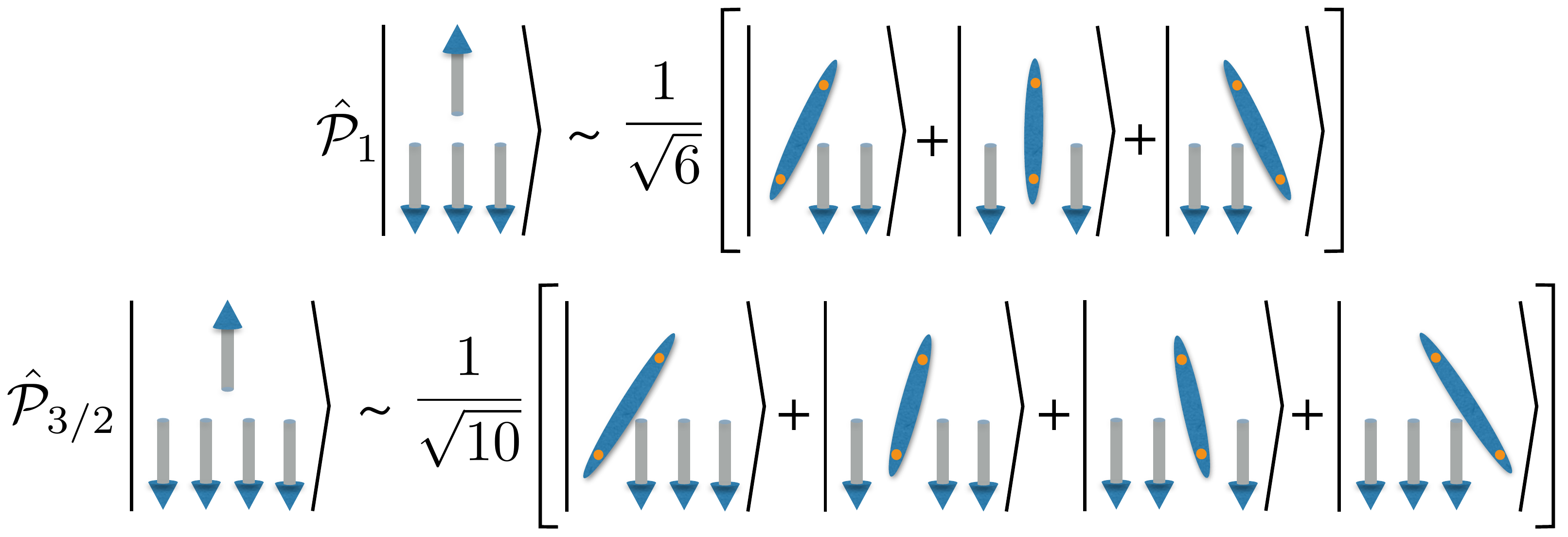} 
\caption{Dark component of $(N+1)$-spin state with one spin up, for $N=3,4$. The dark state is obtained by action of the projection operator $\hat{\mathcal{P}}_{S=(N-1)/2}$, followed by normalization. The result is a linear combination of single-dimer states. The normalization constant for general $N$ is $\sqrt{2/N(N+1)}$.}
\label{fig.one_spin_up}
\end{figure}

The simplest initial state which gives a non-trivial result is  $\vert \!\!\uparrow\downarrow \cdots \downarrow  \rangle$, i.e., one qubit is in the excited state ($M=1$) and $N$ qubits are in the ground state
This gives rise to an RVB state as shown in Fig.~\ref{fig.one_spin_up}. 
The initially excited spin forms a singlet with one of the $N$ available unexcited spins. 
The RVB state is a symmetric linear combination of all such possibilities. This state has `weak' RVB character in that only $2$ out of $N+1$ spins participate in dimer formation. 
This dark state underlies the phenomenon of `radiation trapping'\cite{Stroud1972,Cummings1983}. The probability of null emission is large, $N/(N+1)$, so that very few runs have to discarded in the protocol discussed above.

\section{Initial state with half the spins excited:}
\begin{figure}
\centering
\includegraphics[width=2in]{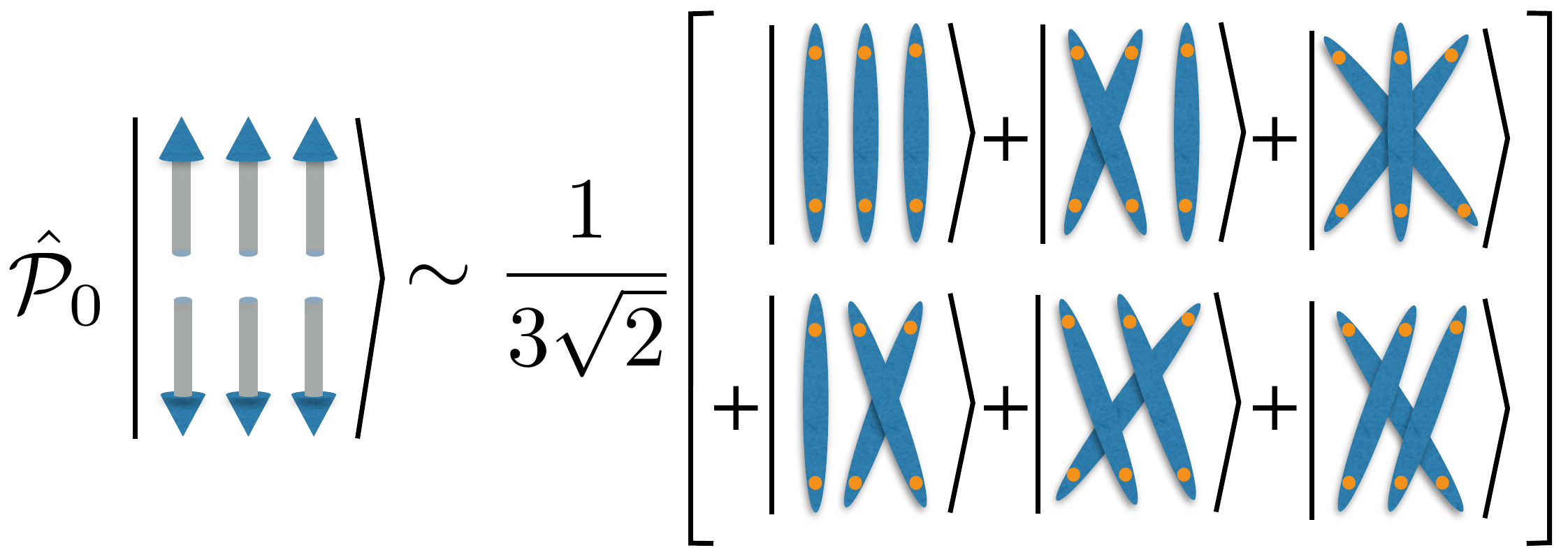}\\
\includegraphics[width=3in]{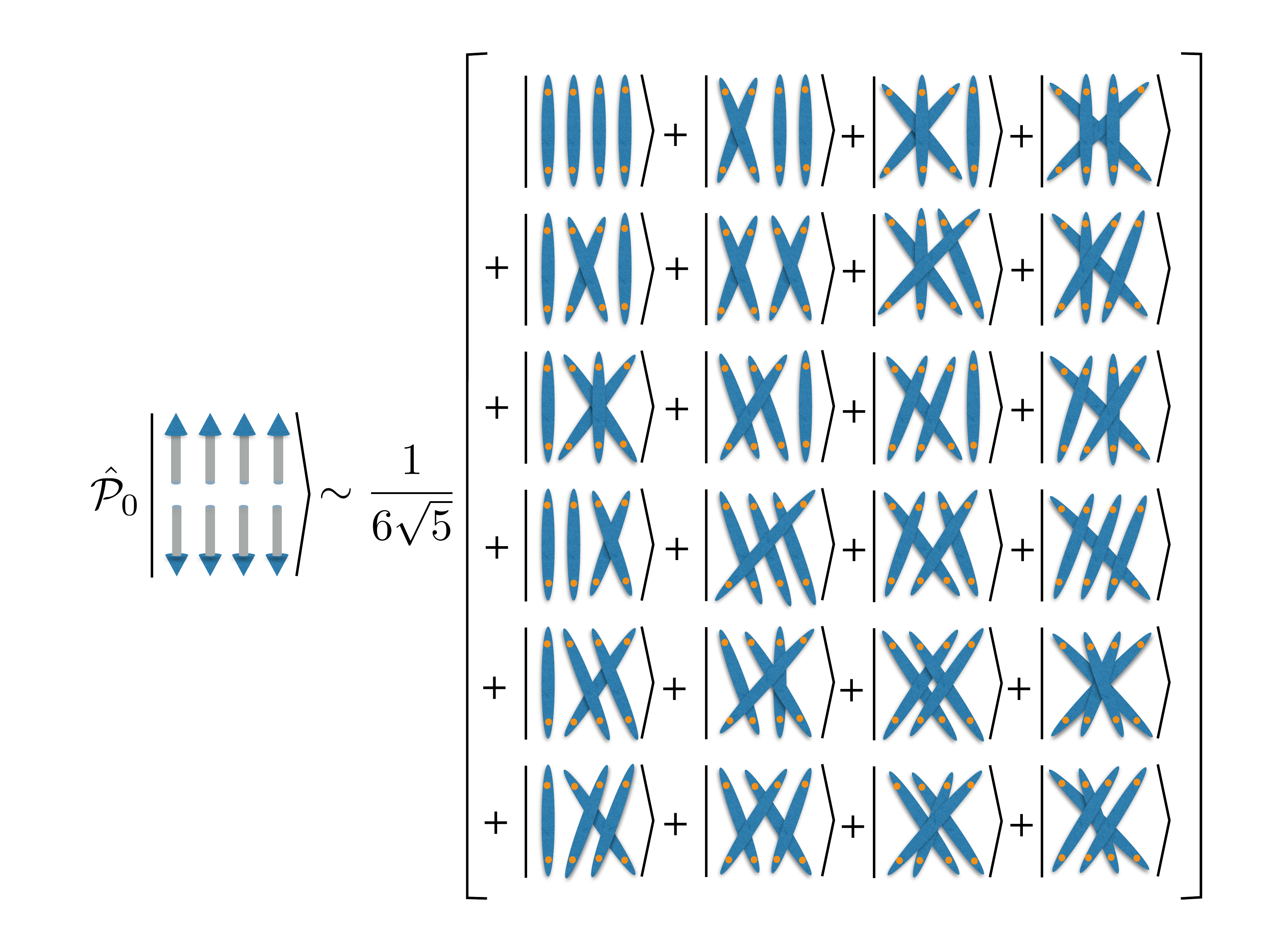}
\caption{Dark component of $(2N)$-spin state with half the spins excited initially, for $N=3,4$. An RVB state results from the action of the singlet projection operator. }
\label{fig.rvb}
\end{figure}

Next, we choose the initial state to be $ \vert \underbrace{\uparrow \ldots \uparrow}_{N}  \underbrace{\downarrow \ldots \downarrow}_{N} \rangle$ with equal number of $\uparrow$ and $\downarrow$ spins ($M=N$). The resulting dark RVB state is shown in Fig.~\ref{fig.rvb} for $N=3,4$. It is a linear combination of $N!$ dimer configurations, which can be seen as follows.
We can initially place $N$ dimers vertically, connecting each spin of the top row with the spin directly below it. We can permute the spins of the bottom row to give rise to different allowed dimer configurations. The number of dimer configurations is thus $N!$, the number of distinct permutations of $N$ spins in the bottom row. 

This RVB state has $S_{tot}=0$; it can neither increase nor decrease its $m_{tot}$ quantum number. It is `doubly dark' as it can neither absorb or emit photons, a key feature  that distinguishes it from other dark states such as that in Fig.~\ref{fig.one_spin_up}. This suggests a clear experimental signature --
when external photons are pumped in, the cavity's transmission characteristics will be the same as that for an empty cavity without spins. This can be used to weed out false positives arising from an imperfect photon detector. 

This state, resulting from the initial state with $N=M$,  has `strong' RVB character in that every spin is part of a dimer. This is an analogue of the RVB ground state proposed for antiferromagnets by Anderson\cite{ANDERSON1973}. 
Tuning $M$ away from $N$ is equivalent to doping, with unpaired spins playing the role of `spinons'. 
Unlike lattice systems where RVB states have been extensively studied (e.g., in Refs.~\onlinecite{Beach2006,Anushya2007,Poilblanc2012}), our system is fully local. Nevertheless, it would be interesting to see if
 signatures of superconductivity emerge. The RVB wavefunction has Slater-permanent character. Thinking of dimers as particles, we assume single-particle states to be indexed by the position within the top row. The position in the bottom row acts as a spatial coordinate. In this language, the RVB state is a Slater permanent -- it resembles the wavefunction of $N$ indistinguishable particles with bosonic character. 

\section{Entanglement in the RVB state:}
The RVB state is manifestly entangled. To quantify this, we notice that  Eq.~\ref{eq.Schmidt_RVB} provides 
a row-wise Schmidt decomposition of the RVB state. We can now partially trace over one of the rows to find its reduced density matrix. For the $N=M$ case, we have
\begin{eqnarray}
\nonumber \vert \Psi_{RVB}^{N=M} \rangle  = \sum_{\lambda=0}^{M}\frac{(-1)^\lambda}{\sqrt{N+1}} \vert S_{tot} = N/2, m_{tot} = N/2-\lambda \rangle_{t} \otimes \\
\vert  S_{tot} = N/2, m_{tot} = \lambda-N/2  \rangle_{b}.\phantom{abcd}
\label{eq.schmidt_b}
\end{eqnarray}
The resulting reduced density matrix for each row is diagonal; the corresponding entanglement entropy is
\begin{equation}
S_{t} = \mathrm{Tr}\{ \hat{\rho}_{top}^r  \log(\hat{\rho}_{top}^r ) \}= \log(N+1),
\end{equation}
which grows with $N$. 
From Eq.~\ref{eq.adark}, the probability for null emission is $1/(N+1)$, falling slowly with $N$. This paves the way to generate entangled states of many particles, e.g., an RVB state of $20$ spins can be created at the cost of discarding $\sim$ 91\% of the runs in the experiment.

\section{Discussion:}
We have shown that a null measurement for photon emission can be used to generate RVB states with high entanglement. 
Our work builds upon earlier studies of measurement-induced entanglement\cite{Chou2005,Moehring2007,Li2008,Julsgaard2012,Roch2014}, subradiance in quantum dots\cite{Sitek2013} and vacuum-induced coherence\cite{agarwal2013quantum,Sitek2012}. These works follow a stochastic approach requiring averaging over several measurements to generate entangled density matrices. In contrast, our proposal exploits wavefunction collapse to give a pure state with entanglement. 
Our scheme also naturally generalises to many-spin systems, producing a specific family of RVB states.

A limiting factor in our proposal is that inter-qubit coupling, e.g. of dipolar character, may cause dephasing. However, interactions may be useful for creating more interesting RVB states, by inducing a preference for shorter range dimers. This is a promising direction for future research as short range RVB states are known to harbour topological order\cite{Kivelson1987}. 

\appendix

\section{Consequences of permutation symmetry}
Here, we expand on some of the mathematical ideas underlying the arguments in the main text. The key ingredient in our discussion is in-row permutation symmetry. This gives rise to three properties which we list as lemmas here.

\begin{itemize}
\item{\underline{Lemma 1}: A state of $M$ spins that is symmetric under any permutation of the $M$ constituent spins must necessarily have $S_{tot}=M/2$, i.e., it has the maximal total angular momentum quantum number. Conversely, if a state of $M$ spins has  $S_{tot}=M/2$, it is symmetric under any permutation of the constituent spins. }

\item{\underline{Lemma 2}: For the system with $M$ spins,  a quantum state with $S_{tot}=M/2$ is uniquely defined by its $m_{tot}$ quantum number, i.e., a state given by $\vert S_{tot} = M/2, m \rangle$ is unique. }

\item{\underline{Lemma 3}: For the system with $M$ spins, the projection operator onto the subspace with fixed $S_{tot}$ quantum number is symmetric under permutations. This can be seen by explicit construction. We have
\begin{equation}
\hat{P}_{S=\Sigma} = \prod_{S\neq \Sigma} \frac{\hat{S}_{tot}^2 - S(S+1)}{\Sigma(\Sigma+1) - S(S+1) }.
\end{equation}
 }
This is easily seen to satisfy all conditions required of a projection operator, e.g., $\hat{P}_{S=\Sigma}^2 = \hat{P}_{S=\Sigma}$ and $\hat{P}_{S=\Sigma}$ acts as the identity operator on any state with $S=\Sigma$. As it only contains the $\hat{S}_{tot}$ operator, it is symmetric under any permutation of the constituent spins. 
\end{itemize}
Lemmas 1 and 2 are known properties of superradiant states. They follow from the fact that $\vert S_{tot} = M/2, m_{tot} = M/2 \rangle = \vert \uparrow \ldots \uparrow \rangle$ is uniquely defined. All other states with $S_{tot} = M/2$ can be obtained by the action of ladder operators on this state, with the ladder operators themselves being permutation symmetric. Lemma 3 is a deep consequence of permutation symmetry, related to Schur-Weyl duality.

In the arguments preceding Eq.~6 in the main text, we use Lemma 3 above to argue that the dark state wavefunction is symmetric under in-row permutations. Subsequently, we perform a row-wise decomposition of the dark state wavefunction. We may write
\begin{eqnarray}
\nonumber \vert \Psi_{dark}^{M,N} \rangle  = \sum_{\mu,\mu'} C_{\mu,\mu'} \vert \mu \rangle_{t} \otimes
\vert  \mu'  \rangle_{b},
\label{eq.rowdecomposition}
\end{eqnarray}
where  $t$ and $b$ denote top and bottom rows respectively. Here, $\mu$ ($\mu'$) are quantum numbers that characterise states from the Hilbert space of the top (bottom) row. 
Now, invoking in-row permutation symmetry of $\vert \Psi_{dark}^{M,N} \rangle $, we conclude that $ \vert \mu \rangle_{t}$ is symmetric under any permutation of the $M$ spins in the top row. Similary, $\vert \mu' \rangle_{b}$ is symmetric under any permutation of the $N$ spins in the bottom row. From Lemma 1, we conclude that $ \vert \mu \rangle_{t}$ has $S_{tot}=M/2$ and $\vert \mu' \rangle_{b}$ has $S_{tot}=N/2$. Furthermore, from Lemma 2, the only quantum number we need to characterise these states is $m_{tot}$. The $m_{tot}$ quantum numbers must be chosen in such a way as to conserve the z-component of angular momentum. This leads to the row-wise decomposition in Eq.~6 of the main text. 

\subsection{Equivalence between dark state and RVB state}
In the arguments after Eq.~10 in the main text, we compare Eq.~6 and Eq.~10. These equations provide row-wise decompositions of $\vert \Psi_{dark}^{M,N}\rangle$ and $\vert \Psi_{RVB}\rangle$. We have explicitly shown that the coefficients in the expansion are equal. To argue that $\vert \Psi_{dark}^{M,N}\rangle$ and $\vert \Psi_{RVB}\rangle$ are identical, we need to ensure that the wavefunctions ($\vert S_{tot=M/2}, m_{tot}={M/2-\kappa} \rangle_{t} $ and $\vert S_{tot=N/2}, m_{tot}={\kappa-N/2} \rangle_{b}$) in the expansion are identical. This follows from Lemma 2 above which ensures that $m_{tot}$ uniquely identifies a state with maximal $S_{tot}$.
\acknowledgments
\section{Acknowledgments:} We thank Amritanshu Prasad, S. Viswanath, Nicol\'as Quesada and Sibasish Ghosh for useful discussions. 
GB thanks the Science and Engineering Research Board (SERB, India) for the SERB Distinguished Fellowship. Research at Perimeter Institute is supported by the Government of Canada through the Department of Innovation,
Science and Economic Development Canada and by the Province of Ontario through the
Ministry of Research, Innovation and Science.

\bibliographystyle{apsrev4-1} 
\bibliography{Dicke_entanglement}

\begin{thebibliography}{41}%
\makeatletter
\providecommand \@ifxundefined [1]{%
 \@ifx{#1\undefined}
}%
\providecommand \@ifnum [1]{%
 \ifnum #1\expandafter \@firstoftwo
 \else \expandafter \@secondoftwo
 \fi
}%
\providecommand \@ifx [1]{%
 \ifx #1\expandafter \@firstoftwo
 \else \expandafter \@secondoftwo
 \fi
}%
\providecommand \natexlab [1]{#1}%
\providecommand \enquote  [1]{``#1''}%
\providecommand \bibnamefont  [1]{#1}%
\providecommand \bibfnamefont [1]{#1}%
\providecommand \citenamefont [1]{#1}%
\providecommand \href@noop [0]{\@secondoftwo}%
\providecommand \href [0]{\begingroup \@sanitize@url \@href}%
\providecommand \@href[1]{\@@startlink{#1}\@@href}%
\providecommand \@@href[1]{\endgroup#1\@@endlink}%
\providecommand \@sanitize@url [0]{\catcode `\\12\catcode `\$12\catcode
  `\&12\catcode `\#12\catcode `\^12\catcode `\_12\catcode `\%12\relax}%
\providecommand \@@startlink[1]{}%
\providecommand \@@endlink[0]{}%
\providecommand \url  [0]{\begingroup\@sanitize@url \@url }%
\providecommand \@url [1]{\endgroup\@href {#1}{\urlprefix }}%
\providecommand \urlprefix  [0]{URL }%
\providecommand \Eprint [0]{\href }%
\providecommand \doibase [0]{http://dx.doi.org/}%
\providecommand \selectlanguage [0]{\@gobble}%
\providecommand \bibinfo  [0]{\@secondoftwo}%
\providecommand \bibfield  [0]{\@secondoftwo}%
\providecommand \translation [1]{[#1]}%
\providecommand \BibitemOpen [0]{}%
\providecommand \bibitemStop [0]{}%
\providecommand \bibitemNoStop [0]{.\EOS\space}%
\providecommand \EOS [0]{\spacefactor3000\relax}%
\providecommand \BibitemShut  [1]{\csname bibitem#1\endcsname}%
\let\auto@bib@innerbib\@empty
\bibitem [{\citenamefont {Dicke}(1954)}]{Dicke1954}%
  \BibitemOpen
  \bibfield  {author} {\bibinfo {author} {\bibfnamefont {R.~H.}\ \bibnamefont
  {Dicke}},\ }\href {\doibase 10.1103/PhysRev.93.99} {\bibfield  {journal}
  {\bibinfo  {journal} {Phys. Rev.}\ }\textbf {\bibinfo {volume} {93}},\
  \bibinfo {pages} {99} (\bibinfo {year} {1954})}\BibitemShut {NoStop}%
\bibitem [{\citenamefont {Inouye}\ \emph {et~al.}(1999)\citenamefont {Inouye},
  \citenamefont {Chikkatur}, \citenamefont {Stamper-Kurn}, \citenamefont
  {Stenger}, \citenamefont {Pritchard},\ and\ \citenamefont
  {Ketterle}}]{Inouye1999}%
  \BibitemOpen
  \bibfield  {author} {\bibinfo {author} {\bibfnamefont {S.}~\bibnamefont
  {Inouye}}, \bibinfo {author} {\bibfnamefont {A.~P.}\ \bibnamefont
  {Chikkatur}}, \bibinfo {author} {\bibfnamefont {D.~M.}\ \bibnamefont
  {Stamper-Kurn}}, \bibinfo {author} {\bibfnamefont {J.}~\bibnamefont
  {Stenger}}, \bibinfo {author} {\bibfnamefont {D.~E.}\ \bibnamefont
  {Pritchard}}, \ and\ \bibinfo {author} {\bibfnamefont {W.}~\bibnamefont
  {Ketterle}},\ }\href {\doibase 10.1126/science.285.5427.571} {\bibfield
  {journal} {\bibinfo  {journal} {Science}\ }\textbf {\bibinfo {volume}
  {285}},\ \bibinfo {pages} {571} (\bibinfo {year} {1999})}\BibitemShut
  {NoStop}%
\bibitem [{\citenamefont {R{\"o}hlsberger}\ \emph {et~al.}(2010)\citenamefont
  {R{\"o}hlsberger}, \citenamefont {Schlage}, \citenamefont {Sahoo},
  \citenamefont {Couet},\ and\ \citenamefont {R{\"u}ffer}}]{Lambshift2010}%
  \BibitemOpen
  \bibfield  {author} {\bibinfo {author} {\bibfnamefont {R.}~\bibnamefont
  {R{\"o}hlsberger}}, \bibinfo {author} {\bibfnamefont {K.}~\bibnamefont
  {Schlage}}, \bibinfo {author} {\bibfnamefont {B.}~\bibnamefont {Sahoo}},
  \bibinfo {author} {\bibfnamefont {S.}~\bibnamefont {Couet}}, \ and\ \bibinfo
  {author} {\bibfnamefont {R.}~\bibnamefont {R{\"u}ffer}},\ }\href {\doibase
  10.1126/science.1187770} {\bibfield  {journal} {\bibinfo  {journal}
  {Science}\ }\textbf {\bibinfo {volume} {328}},\ \bibinfo {pages} {1248}
  (\bibinfo {year} {2010})}\BibitemShut {NoStop}%
\bibitem [{\citenamefont {Rezende}(2009)}]{Rezende2009}%
  \BibitemOpen
  \bibfield  {author} {\bibinfo {author} {\bibfnamefont {S.~M.}\ \bibnamefont
  {Rezende}},\ }\href {\doibase 10.1103/PhysRevB.79.060410} {\bibfield
  {journal} {\bibinfo  {journal} {Phys. Rev. B}\ }\textbf {\bibinfo {volume}
  {79}},\ \bibinfo {pages} {060410} (\bibinfo {year} {2009})}\BibitemShut
  {NoStop}%
\bibitem [{\citenamefont {Tighineanu}\ \emph {et~al.}(2016)\citenamefont
  {Tighineanu}, \citenamefont {Daveau}, \citenamefont {Lehmann}, \citenamefont
  {Beere}, \citenamefont {Ritchie}, \citenamefont {Lodahl},\ and\ \citenamefont
  {Stobbe}}]{QuantumDot}%
  \BibitemOpen
  \bibfield  {author} {\bibinfo {author} {\bibfnamefont {P.}~\bibnamefont
  {Tighineanu}}, \bibinfo {author} {\bibfnamefont {R.~S.}\ \bibnamefont
  {Daveau}}, \bibinfo {author} {\bibfnamefont {T.~B.}\ \bibnamefont {Lehmann}},
  \bibinfo {author} {\bibfnamefont {H.~E.}\ \bibnamefont {Beere}}, \bibinfo
  {author} {\bibfnamefont {D.~A.}\ \bibnamefont {Ritchie}}, \bibinfo {author}
  {\bibfnamefont {P.}~\bibnamefont {Lodahl}}, \ and\ \bibinfo {author}
  {\bibfnamefont {S.}~\bibnamefont {Stobbe}},\ }\href {\doibase
  10.1103/PhysRevLett.116.163604} {\bibfield  {journal} {\bibinfo  {journal}
  {Phys. Rev. Lett.}\ }\textbf {\bibinfo {volume} {116}},\ \bibinfo {pages}
  {163604} (\bibinfo {year} {2016})}\BibitemShut {NoStop}%
\bibitem [{\citenamefont {Mlynek}\ \emph {et~al.}(2014)\citenamefont {Mlynek},
  \citenamefont {Abdumalikov}, \citenamefont {Eichler},\ and\ \citenamefont
  {Wallraff}}]{Mlynek2014}%
  \BibitemOpen
  \bibfield  {author} {\bibinfo {author} {\bibfnamefont {J.~A.}\ \bibnamefont
  {Mlynek}}, \bibinfo {author} {\bibfnamefont {A.~A.}\ \bibnamefont
  {Abdumalikov}}, \bibinfo {author} {\bibfnamefont {C.}~\bibnamefont
  {Eichler}}, \ and\ \bibinfo {author} {\bibfnamefont {A.}~\bibnamefont
  {Wallraff}},\ }\href {http://dx.doi.org/10.1038/ncomms6186} {\bibfield
  {journal} {\bibinfo  {journal} {Nat. Commun.}\ }\textbf {\bibinfo {volume}
  {5}} (\bibinfo {year} {2014})}\BibitemShut {NoStop}%
\bibitem [{\citenamefont {Blais}\ \emph {et~al.}(2004)\citenamefont {Blais},
  \citenamefont {Huang}, \citenamefont {Wallraff}, \citenamefont {Girvin},\
  and\ \citenamefont {Schoelkopf}}]{Blais2004}%
  \BibitemOpen
  \bibfield  {author} {\bibinfo {author} {\bibfnamefont {A.}~\bibnamefont
  {Blais}}, \bibinfo {author} {\bibfnamefont {R.-S.}\ \bibnamefont {Huang}},
  \bibinfo {author} {\bibfnamefont {A.}~\bibnamefont {Wallraff}}, \bibinfo
  {author} {\bibfnamefont {S.~M.}\ \bibnamefont {Girvin}}, \ and\ \bibinfo
  {author} {\bibfnamefont {R.~J.}\ \bibnamefont {Schoelkopf}},\ }\href
  {\doibase 10.1103/PhysRevA.69.062320} {\bibfield  {journal} {\bibinfo
  {journal} {Phys. Rev. A}\ }\textbf {\bibinfo {volume} {69}},\ \bibinfo
  {pages} {062320} (\bibinfo {year} {2004})}\BibitemShut {NoStop}%
\bibitem [{\citenamefont {Ou}\ and\ \citenamefont {Mandel}(1988)}]{Ou1988}%
  \BibitemOpen
  \bibfield  {author} {\bibinfo {author} {\bibfnamefont {Z.~Y.}\ \bibnamefont
  {Ou}}\ and\ \bibinfo {author} {\bibfnamefont {L.}~\bibnamefont {Mandel}},\
  }\href {\doibase 10.1103/PhysRevLett.61.50} {\bibfield  {journal} {\bibinfo
  {journal} {Phys. Rev. Lett.}\ }\textbf {\bibinfo {volume} {61}},\ \bibinfo
  {pages} {50} (\bibinfo {year} {1988})}\BibitemShut {NoStop}%
\bibitem [{\citenamefont {Gao}\ \emph {et~al.}(2010)\citenamefont {Gao},
  \citenamefont {Lu}, \citenamefont {Yao}, \citenamefont {Xu}, \citenamefont
  {Guhne}, \citenamefont {Goebel}, \citenamefont {Chen}, \citenamefont {Peng},
  \citenamefont {Chen},\ and\ \citenamefont {Pan}}]{Gao2010}%
  \BibitemOpen
  \bibfield  {author} {\bibinfo {author} {\bibfnamefont {W.-B.}\ \bibnamefont
  {Gao}}, \bibinfo {author} {\bibfnamefont {C.-Y.}\ \bibnamefont {Lu}},
  \bibinfo {author} {\bibfnamefont {X.-C.}\ \bibnamefont {Yao}}, \bibinfo
  {author} {\bibfnamefont {P.}~\bibnamefont {Xu}}, \bibinfo {author}
  {\bibfnamefont {O.}~\bibnamefont {Guhne}}, \bibinfo {author} {\bibfnamefont
  {A.}~\bibnamefont {Goebel}}, \bibinfo {author} {\bibfnamefont {Y.-A.}\
  \bibnamefont {Chen}}, \bibinfo {author} {\bibfnamefont {C.-Z.}\ \bibnamefont
  {Peng}}, \bibinfo {author} {\bibfnamefont {Z.-B.}\ \bibnamefont {Chen}}, \
  and\ \bibinfo {author} {\bibfnamefont {J.-W.}\ \bibnamefont {Pan}},\ }\href
  {http://dx.doi.org/10.1038/nphys1603} {\bibfield  {journal} {\bibinfo
  {journal} {Nat Phys}\ }\textbf {\bibinfo {volume} {6}},\ \bibinfo {pages}
  {331} (\bibinfo {year} {2010})}\BibitemShut {NoStop}%
\bibitem [{\citenamefont {Monz}\ \emph {et~al.}(2011)\citenamefont {Monz},
  \citenamefont {Schindler}, \citenamefont {Barreiro}, \citenamefont {Chwalla},
  \citenamefont {Nigg}, \citenamefont {Coish}, \citenamefont {Harlander},
  \citenamefont {H\"ansel}, \citenamefont {Hennrich},\ and\ \citenamefont
  {Blatt}}]{Monz2011}%
  \BibitemOpen
  \bibfield  {author} {\bibinfo {author} {\bibfnamefont {T.}~\bibnamefont
  {Monz}}, \bibinfo {author} {\bibfnamefont {P.}~\bibnamefont {Schindler}},
  \bibinfo {author} {\bibfnamefont {J.~T.}\ \bibnamefont {Barreiro}}, \bibinfo
  {author} {\bibfnamefont {M.}~\bibnamefont {Chwalla}}, \bibinfo {author}
  {\bibfnamefont {D.}~\bibnamefont {Nigg}}, \bibinfo {author} {\bibfnamefont
  {W.~A.}\ \bibnamefont {Coish}}, \bibinfo {author} {\bibfnamefont
  {M.}~\bibnamefont {Harlander}}, \bibinfo {author} {\bibfnamefont
  {W.}~\bibnamefont {H\"ansel}}, \bibinfo {author} {\bibfnamefont
  {M.}~\bibnamefont {Hennrich}}, \ and\ \bibinfo {author} {\bibfnamefont
  {R.}~\bibnamefont {Blatt}},\ }\href {\doibase 10.1103/PhysRevLett.106.130506}
  {\bibfield  {journal} {\bibinfo  {journal} {Phys. Rev. Lett.}\ }\textbf
  {\bibinfo {volume} {106}},\ \bibinfo {pages} {130506} (\bibinfo {year}
  {2011})}\BibitemShut {NoStop}%
\bibitem [{\citenamefont {{Wang}}\ \emph {et~al.}(2016)\citenamefont {{Wang}},
  \citenamefont {{Chen}}, \citenamefont {{Li}}, \citenamefont {{Huang}},
  \citenamefont {{Liu}}, \citenamefont {{Chen}}, \citenamefont {{Luo}},
  \citenamefont {{Su}}, \citenamefont {{Wu}}, \citenamefont {{Li}},
  \citenamefont {{Lu}}, \citenamefont {{Hu}}, \citenamefont {{Jiang}},
  \citenamefont {{Peng}}, \citenamefont {{Li}}, \citenamefont {{Liu}},
  \citenamefont {{Chen}}, \citenamefont {{Lu}},\ and\ \citenamefont
  {{Pan}}}]{Wang2016}%
  \BibitemOpen
  \bibfield  {author} {\bibinfo {author} {\bibfnamefont {X.-L.}\ \bibnamefont
  {{Wang}}}, \bibinfo {author} {\bibfnamefont {L.-K.}\ \bibnamefont {{Chen}}},
  \bibinfo {author} {\bibfnamefont {W.}~\bibnamefont {{Li}}}, \bibinfo {author}
  {\bibfnamefont {H.-L.}\ \bibnamefont {{Huang}}}, \bibinfo {author}
  {\bibfnamefont {C.}~\bibnamefont {{Liu}}}, \bibinfo {author} {\bibfnamefont
  {C.}~\bibnamefont {{Chen}}}, \bibinfo {author} {\bibfnamefont {Y.-H.}\
  \bibnamefont {{Luo}}}, \bibinfo {author} {\bibfnamefont {Z.-E.}\ \bibnamefont
  {{Su}}}, \bibinfo {author} {\bibfnamefont {D.}~\bibnamefont {{Wu}}}, \bibinfo
  {author} {\bibfnamefont {Z.-D.}\ \bibnamefont {{Li}}}, \bibinfo {author}
  {\bibfnamefont {H.}~\bibnamefont {{Lu}}}, \bibinfo {author} {\bibfnamefont
  {Y.}~\bibnamefont {{Hu}}}, \bibinfo {author} {\bibfnamefont {X.}~\bibnamefont
  {{Jiang}}}, \bibinfo {author} {\bibfnamefont {C.-Z.}\ \bibnamefont {{Peng}}},
  \bibinfo {author} {\bibfnamefont {L.}~\bibnamefont {{Li}}}, \bibinfo {author}
  {\bibfnamefont {N.-L.}\ \bibnamefont {{Liu}}}, \bibinfo {author}
  {\bibfnamefont {Y.-A.}\ \bibnamefont {{Chen}}}, \bibinfo {author}
  {\bibfnamefont {C.-Y.}\ \bibnamefont {{Lu}}}, \ and\ \bibinfo {author}
  {\bibfnamefont {J.-W.}\ \bibnamefont {{Pan}}},\ }\href@noop {} {\bibfield
  {journal} {\bibinfo  {journal} {ArXiv e-prints}\ } (\bibinfo {year}
  {2016})},\ \Eprint {http://arxiv.org/abs/1605.08547} {arXiv:1605.08547
  [quant-ph]} \BibitemShut {NoStop}%
\bibitem [{\citenamefont {H\"ark\"onen}\ \emph {et~al.}(2009)\citenamefont
  {H\"ark\"onen}, \citenamefont {Plastina},\ and\ \citenamefont
  {Maniscalco}}]{Harkonen2009}%
  \BibitemOpen
  \bibfield  {author} {\bibinfo {author} {\bibfnamefont {K.}~\bibnamefont
  {H\"ark\"onen}}, \bibinfo {author} {\bibfnamefont {F.}~\bibnamefont
  {Plastina}}, \ and\ \bibinfo {author} {\bibfnamefont {S.}~\bibnamefont
  {Maniscalco}},\ }\href {\doibase 10.1103/PhysRevA.80.033841} {\bibfield
  {journal} {\bibinfo  {journal} {Phys. Rev. A}\ }\textbf {\bibinfo {volume}
  {80}},\ \bibinfo {pages} {033841} (\bibinfo {year} {2009})}\BibitemShut
  {NoStop}%
\bibitem [{\citenamefont {Francica}\ \emph {et~al.}(2009)\citenamefont
  {Francica}, \citenamefont {Maniscalco}, \citenamefont {Piilo}, \citenamefont
  {Plastina},\ and\ \citenamefont {Suominen}}]{Maniscalco2009}%
  \BibitemOpen
  \bibfield  {author} {\bibinfo {author} {\bibfnamefont {F.}~\bibnamefont
  {Francica}}, \bibinfo {author} {\bibfnamefont {S.}~\bibnamefont
  {Maniscalco}}, \bibinfo {author} {\bibfnamefont {J.}~\bibnamefont {Piilo}},
  \bibinfo {author} {\bibfnamefont {F.}~\bibnamefont {Plastina}}, \ and\
  \bibinfo {author} {\bibfnamefont {K.-A.}\ \bibnamefont {Suominen}},\ }\href
  {\doibase 10.1103/PhysRevA.79.032310} {\bibfield  {journal} {\bibinfo
  {journal} {Phys. Rev. A}\ }\textbf {\bibinfo {volume} {79}},\ \bibinfo
  {pages} {032310} (\bibinfo {year} {2009})}\BibitemShut {NoStop}%
\bibitem [{\citenamefont {Wolfe}\ and\ \citenamefont
  {Yelin}(2014)}]{Wolfe2014}%
  \BibitemOpen
  \bibfield  {author} {\bibinfo {author} {\bibfnamefont {E.}~\bibnamefont
  {Wolfe}}\ and\ \bibinfo {author} {\bibfnamefont {S.~F.}\ \bibnamefont
  {Yelin}},\ }\href {\doibase 10.1103/PhysRevLett.112.140402} {\bibfield
  {journal} {\bibinfo  {journal} {Phys. Rev. Lett.}\ }\textbf {\bibinfo
  {volume} {112}},\ \bibinfo {pages} {140402} (\bibinfo {year}
  {2014})}\BibitemShut {NoStop}%
\bibitem [{\citenamefont {Scully}(2015)}]{Scully2015}%
  \BibitemOpen
  \bibfield  {author} {\bibinfo {author} {\bibfnamefont {M.~O.}\ \bibnamefont
  {Scully}},\ }\href {\doibase 10.1103/PhysRevLett.115.243602} {\bibfield
  {journal} {\bibinfo  {journal} {Phys. Rev. Lett.}\ }\textbf {\bibinfo
  {volume} {115}},\ \bibinfo {pages} {243602} (\bibinfo {year}
  {2015})}\BibitemShut {NoStop}%
\bibitem [{\citenamefont {{McGuyer}}\ \emph {et~al.}(2015)\citenamefont
  {{McGuyer}}, \citenamefont {{McDonald}}, \citenamefont {{Iwata}},
  \citenamefont {{Tarallo}}, \citenamefont {{Skomorowski}}, \citenamefont
  {{Moszynski}},\ and\ \citenamefont {{Zelevinsky}}}]{McGuyer2015}%
  \BibitemOpen
  \bibfield  {author} {\bibinfo {author} {\bibfnamefont {B.~H.}\ \bibnamefont
  {{McGuyer}}}, \bibinfo {author} {\bibfnamefont {M.}~\bibnamefont
  {{McDonald}}}, \bibinfo {author} {\bibfnamefont {G.~Z.}\ \bibnamefont
  {{Iwata}}}, \bibinfo {author} {\bibfnamefont {M.~G.}\ \bibnamefont
  {{Tarallo}}}, \bibinfo {author} {\bibfnamefont {W.}~\bibnamefont
  {{Skomorowski}}}, \bibinfo {author} {\bibfnamefont {R.}~\bibnamefont
  {{Moszynski}}}, \ and\ \bibinfo {author} {\bibfnamefont {T.}~\bibnamefont
  {{Zelevinsky}}},\ }\href {\doibase 10.1038/nphys3182} {\bibfield  {journal}
  {\bibinfo  {journal} {Nature Physics}\ }\textbf {\bibinfo {volume} {11}},\
  \bibinfo {pages} {32} (\bibinfo {year} {2015})},\ \Eprint
  {http://arxiv.org/abs/1407.4752} {arXiv:1407.4752 [physics.atom-ph]}
  \BibitemShut {NoStop}%
\bibitem [{\citenamefont {Mirza}\ and\ \citenamefont
  {Begzjav}(2016)}]{Mirza2016}%
  \BibitemOpen
  \bibfield  {author} {\bibinfo {author} {\bibfnamefont {I.~M.}\ \bibnamefont
  {Mirza}}\ and\ \bibinfo {author} {\bibfnamefont {T.}~\bibnamefont
  {Begzjav}},\ }\href {http://stacks.iop.org/0295-5075/114/i=2/a=24004}
  {\bibfield  {journal} {\bibinfo  {journal} {EPL (Europhysics Letters)}\
  }\textbf {\bibinfo {volume} {114}},\ \bibinfo {pages} {24004} (\bibinfo
  {year} {2016})}\BibitemShut {NoStop}%
\bibitem [{\citenamefont {Pauling}(1960)}]{Pauling1960}%
  \BibitemOpen
  \bibfield  {author} {\bibinfo {author} {\bibfnamefont {L.}~\bibnamefont
  {Pauling}},\ }\href {https://books.google.co.in/books?id=L-1K9HmKmUUC} {\emph
  {\bibinfo {title} {The Nature of the Chemical Bond
  Molecules and Crystals: An Introduction to Modern Structural Chemistry}}},\
  George Fisher Baker Non-Resident Lecture Series\ (\bibinfo  {publisher}
  {Cornell University Press},\ \bibinfo {year} {1960})\BibitemShut {NoStop}%
\bibitem [{\citenamefont {Anderson}(1973)}]{ANDERSON1973}%
  \BibitemOpen
  \bibfield  {author} {\bibinfo {author} {\bibfnamefont {P.}~\bibnamefont
  {Anderson}},\ }\href {\doibase
  http://dx.doi.org/10.1016/0025-5408(73)90167-0} {\bibfield  {journal}
  {\bibinfo  {journal} {Materials Research Bulletin}\ }\textbf {\bibinfo
  {volume} {8}},\ \bibinfo {pages} {153 } (\bibinfo {year} {1973})}\BibitemShut
  {NoStop}%
\bibitem [{\citenamefont {Baskaran}\ \emph {et~al.}(1987)\citenamefont
  {Baskaran}, \citenamefont {Zou},\ and\ \citenamefont {Anderson}}]{BZA1987}%
  \BibitemOpen
  \bibfield  {author} {\bibinfo {author} {\bibfnamefont {G.}~\bibnamefont
  {Baskaran}}, \bibinfo {author} {\bibfnamefont {Z.}~\bibnamefont {Zou}}, \
  and\ \bibinfo {author} {\bibfnamefont {P.}~\bibnamefont {Anderson}},\ }\href
  {\doibase http://dx.doi.org/10.1016/0038-1098(87)90642-9} {\bibfield
  {journal} {\bibinfo  {journal} {Solid State Communications}\ }\textbf
  {\bibinfo {volume} {63}},\ \bibinfo {pages} {973 } (\bibinfo {year}
  {1987})}\BibitemShut {NoStop}%
\bibitem [{\citenamefont {Kivelson}\ \emph {et~al.}(1987)\citenamefont
  {Kivelson}, \citenamefont {Rokhsar},\ and\ \citenamefont
  {Sethna}}]{Kivelson1987}%
  \BibitemOpen
  \bibfield  {author} {\bibinfo {author} {\bibfnamefont {S.~A.}\ \bibnamefont
  {Kivelson}}, \bibinfo {author} {\bibfnamefont {D.~S.}\ \bibnamefont
  {Rokhsar}}, \ and\ \bibinfo {author} {\bibfnamefont {J.~P.}\ \bibnamefont
  {Sethna}},\ }\href {\doibase 10.1103/PhysRevB.35.8865} {\bibfield  {journal}
  {\bibinfo  {journal} {Phys. Rev. B}\ }\textbf {\bibinfo {volume} {35}},\
  \bibinfo {pages} {8865} (\bibinfo {year} {1987})}\BibitemShut {NoStop}%
\bibitem [{\citenamefont {Trebst}\ \emph {et~al.}(2006)\citenamefont {Trebst},
  \citenamefont {Schollw\"ock}, \citenamefont {Troyer},\ and\ \citenamefont
  {Zoller}}]{Trebst2006}%
  \BibitemOpen
  \bibfield  {author} {\bibinfo {author} {\bibfnamefont {S.}~\bibnamefont
  {Trebst}}, \bibinfo {author} {\bibfnamefont {U.}~\bibnamefont
  {Schollw\"ock}}, \bibinfo {author} {\bibfnamefont {M.}~\bibnamefont
  {Troyer}}, \ and\ \bibinfo {author} {\bibfnamefont {P.}~\bibnamefont
  {Zoller}},\ }\href {\doibase 10.1103/PhysRevLett.96.250402} {\bibfield
  {journal} {\bibinfo  {journal} {Phys. Rev. Lett.}\ }\textbf {\bibinfo
  {volume} {96}},\ \bibinfo {pages} {250402} (\bibinfo {year}
  {2006})}\BibitemShut {NoStop}%
\bibitem [{\citenamefont {Nascimb\`ene}\ \emph {et~al.}(2012)\citenamefont
  {Nascimb\`ene}, \citenamefont {Chen}, \citenamefont {Atala}, \citenamefont
  {Aidelsburger}, \citenamefont {Trotzky}, \citenamefont {Paredes},\ and\
  \citenamefont {Bloch}}]{Nascimbene2012}%
  \BibitemOpen
  \bibfield  {author} {\bibinfo {author} {\bibfnamefont {S.}~\bibnamefont
  {Nascimb\`ene}}, \bibinfo {author} {\bibfnamefont {Y.-A.}\ \bibnamefont
  {Chen}}, \bibinfo {author} {\bibfnamefont {M.}~\bibnamefont {Atala}},
  \bibinfo {author} {\bibfnamefont {M.}~\bibnamefont {Aidelsburger}}, \bibinfo
  {author} {\bibfnamefont {S.}~\bibnamefont {Trotzky}}, \bibinfo {author}
  {\bibfnamefont {B.}~\bibnamefont {Paredes}}, \ and\ \bibinfo {author}
  {\bibfnamefont {I.}~\bibnamefont {Bloch}},\ }\href {\doibase
  10.1103/PhysRevLett.108.205301} {\bibfield  {journal} {\bibinfo  {journal}
  {Phys. Rev. Lett.}\ }\textbf {\bibinfo {volume} {108}},\ \bibinfo {pages}
  {205301} (\bibinfo {year} {2012})}\BibitemShut {NoStop}%
\bibitem [{\citenamefont {Tavis}\ and\ \citenamefont
  {Cummings}(1968)}]{Tavis1968}%
  \BibitemOpen
  \bibfield  {author} {\bibinfo {author} {\bibfnamefont {M.}~\bibnamefont
  {Tavis}}\ and\ \bibinfo {author} {\bibfnamefont {F.~W.}\ \bibnamefont
  {Cummings}},\ }\href {\doibase 10.1103/PhysRev.170.379} {\bibfield  {journal}
  {\bibinfo  {journal} {Phys. Rev.}\ }\textbf {\bibinfo {volume} {170}},\
  \bibinfo {pages} {379} (\bibinfo {year} {1968})}\BibitemShut {NoStop}%
\bibitem [{\citenamefont {Jaynes}\ and\ \citenamefont
  {Cummings}(1963)}]{Jaynes1963}%
  \BibitemOpen
  \bibfield  {author} {\bibinfo {author} {\bibfnamefont {E.~T.}\ \bibnamefont
  {Jaynes}}\ and\ \bibinfo {author} {\bibfnamefont {F.~W.}\ \bibnamefont
  {Cummings}},\ }\href {\doibase 10.1109/PROC.1963.1664} {\bibfield  {journal}
  {\bibinfo  {journal} {Proceedings of the IEEE}\ }\textbf {\bibinfo {volume}
  {51}},\ \bibinfo {pages} {89} (\bibinfo {year} {1963})}\BibitemShut {NoStop}%
\bibitem [{\citenamefont {Klimov}\ and\ \citenamefont
  {Chumakov}(2009)}]{klimov2009group}%
  \BibitemOpen
  \bibfield  {author} {\bibinfo {author} {\bibfnamefont {A.}~\bibnamefont
  {Klimov}}\ and\ \bibinfo {author} {\bibfnamefont {S.}~\bibnamefont
  {Chumakov}},\ }\href {https://books.google.co.in/books?id=ivmAxQ0G9oUC}
  {\emph {\bibinfo {title} {A Group-Theoretical Approach to Quantum Optics}}}\
  (\bibinfo  {publisher} {Wiley},\ \bibinfo {year} {2009})\BibitemShut
  {NoStop}%
\bibitem [{\citenamefont {Rose}(1957)}]{Rose}%
  \BibitemOpen
  \bibfield  {author} {\bibinfo {author} {\bibfnamefont {M.}~\bibnamefont
  {Rose}},\ }\href {https://books.google.ca/books?id=QApRAAAAMAAJ} {\emph
  {\bibinfo {title} {Elementary theory of angular momentum}}},\ Structure of
  matter series\ (\bibinfo  {publisher} {Wiley},\ \bibinfo {year}
  {1957})\BibitemShut {NoStop}%
\bibitem [{\citenamefont {Marshall}(1955)}]{Marshall1955}%
  \BibitemOpen
  \bibfield  {author} {\bibinfo {author} {\bibfnamefont {W.}~\bibnamefont
  {Marshall}},\ }\href {\doibase 10.1098/rspa.1955.0200} {\bibfield  {journal}
  {\bibinfo  {journal} {Proceedings of the Royal Society of London A:
  Mathematical, Physical and Engineering Sciences}\ }\textbf {\bibinfo {volume}
  {232}},\ \bibinfo {pages} {48} (\bibinfo {year} {1955})},\ \Eprint
  {http://arxiv.org/abs/http://rspa.royalsocietypublishing.org/content/232/1188/48.full.pdf}
  {http://rspa.royalsocietypublishing.org/content/232/1188/48.full.pdf}
  \BibitemShut {NoStop}%
\bibitem [{\citenamefont {Stroud}\ \emph {et~al.}(1972)\citenamefont {Stroud},
  \citenamefont {Eberly}, \citenamefont {Lama},\ and\ \citenamefont
  {Mandel}}]{Stroud1972}%
  \BibitemOpen
  \bibfield  {author} {\bibinfo {author} {\bibfnamefont {C.~R.}\ \bibnamefont
  {Stroud}}, \bibinfo {author} {\bibfnamefont {J.~H.}\ \bibnamefont {Eberly}},
  \bibinfo {author} {\bibfnamefont {W.~L.}\ \bibnamefont {Lama}}, \ and\
  \bibinfo {author} {\bibfnamefont {L.}~\bibnamefont {Mandel}},\ }\href
  {\doibase 10.1103/PhysRevA.5.1094} {\bibfield  {journal} {\bibinfo  {journal}
  {Phys. Rev. A}\ }\textbf {\bibinfo {volume} {5}},\ \bibinfo {pages} {1094}
  (\bibinfo {year} {1972})}\BibitemShut {NoStop}%
\bibitem [{\citenamefont {Cummings}\ and\ \citenamefont
  {Dorri}(1983)}]{Cummings1983}%
  \BibitemOpen
  \bibfield  {author} {\bibinfo {author} {\bibfnamefont {F.~W.}\ \bibnamefont
  {Cummings}}\ and\ \bibinfo {author} {\bibfnamefont {A.}~\bibnamefont
  {Dorri}},\ }\href {\doibase 10.1103/PhysRevA.28.2282} {\bibfield  {journal}
  {\bibinfo  {journal} {Phys. Rev. A}\ }\textbf {\bibinfo {volume} {28}},\
  \bibinfo {pages} {2282} (\bibinfo {year} {1983})}\BibitemShut {NoStop}%
\bibitem [{\citenamefont {Beach}\ and\ \citenamefont
  {Sandvik}(2006)}]{Beach2006}%
  \BibitemOpen
  \bibfield  {author} {\bibinfo {author} {\bibfnamefont {K.}~\bibnamefont
  {Beach}}\ and\ \bibinfo {author} {\bibfnamefont {A.~W.}\ \bibnamefont
  {Sandvik}},\ }\href {\doibase
  http://dx.doi.org/10.1016/j.nuclphysb.2006.05.032} {\bibfield  {journal}
  {\bibinfo  {journal} {Nuclear Physics B}\ }\textbf {\bibinfo {volume}
  {750}},\ \bibinfo {pages} {142 } (\bibinfo {year} {2006})}\BibitemShut
  {NoStop}%
\bibitem [{\citenamefont {Chandran}\ \emph {et~al.}(2007)\citenamefont
  {Chandran}, \citenamefont {Kaszlikowski}, \citenamefont {Sen(De)},
  \citenamefont {Sen},\ and\ \citenamefont {Vedral}}]{Anushya2007}%
  \BibitemOpen
  \bibfield  {author} {\bibinfo {author} {\bibfnamefont {A.}~\bibnamefont
  {Chandran}}, \bibinfo {author} {\bibfnamefont {D.}~\bibnamefont
  {Kaszlikowski}}, \bibinfo {author} {\bibfnamefont {A.}~\bibnamefont
  {Sen(De)}}, \bibinfo {author} {\bibfnamefont {U.}~\bibnamefont {Sen}}, \ and\
  \bibinfo {author} {\bibfnamefont {V.}~\bibnamefont {Vedral}},\ }\href
  {\doibase 10.1103/PhysRevLett.99.170502} {\bibfield  {journal} {\bibinfo
  {journal} {Phys. Rev. Lett.}\ }\textbf {\bibinfo {volume} {99}},\ \bibinfo
  {pages} {170502} (\bibinfo {year} {2007})}\BibitemShut {NoStop}%
\bibitem [{\citenamefont {Poilblanc}\ \emph {et~al.}(2012)\citenamefont
  {Poilblanc}, \citenamefont {Schuch}, \citenamefont {P\'erez-Garc\'{\i}a},\
  and\ \citenamefont {Cirac}}]{Poilblanc2012}%
  \BibitemOpen
  \bibfield  {author} {\bibinfo {author} {\bibfnamefont {D.}~\bibnamefont
  {Poilblanc}}, \bibinfo {author} {\bibfnamefont {N.}~\bibnamefont {Schuch}},
  \bibinfo {author} {\bibfnamefont {D.}~\bibnamefont {P\'erez-Garc\'{\i}a}}, \
  and\ \bibinfo {author} {\bibfnamefont {J.~I.}\ \bibnamefont {Cirac}},\ }\href
  {\doibase 10.1103/PhysRevB.86.014404} {\bibfield  {journal} {\bibinfo
  {journal} {Phys. Rev. B}\ }\textbf {\bibinfo {volume} {86}},\ \bibinfo
  {pages} {014404} (\bibinfo {year} {2012})}\BibitemShut {NoStop}%
\bibitem [{\citenamefont {Chou}\ \emph {et~al.}(2005)\citenamefont {Chou},
  \citenamefont {de~Riedmatten}, \citenamefont {Felinto}, \citenamefont
  {Polyakov}, \citenamefont {van Enk},\ and\ \citenamefont
  {Kimble}}]{Chou2005}%
  \BibitemOpen
  \bibfield  {author} {\bibinfo {author} {\bibfnamefont {C.~W.}\ \bibnamefont
  {Chou}}, \bibinfo {author} {\bibfnamefont {H.}~\bibnamefont {de~Riedmatten}},
  \bibinfo {author} {\bibfnamefont {D.}~\bibnamefont {Felinto}}, \bibinfo
  {author} {\bibfnamefont {S.~V.}\ \bibnamefont {Polyakov}}, \bibinfo {author}
  {\bibfnamefont {S.~J.}\ \bibnamefont {van Enk}}, \ and\ \bibinfo {author}
  {\bibfnamefont {H.~J.}\ \bibnamefont {Kimble}},\ }\href
  {http://dx.doi.org/10.1038/nature04353} {\bibfield  {journal} {\bibinfo
  {journal} {Nature}\ }\textbf {\bibinfo {volume} {438}},\ \bibinfo {pages}
  {828} (\bibinfo {year} {2005})}\BibitemShut {NoStop}%
\bibitem [{\citenamefont {Moehring}\ \emph {et~al.}(2007)\citenamefont
  {Moehring}, \citenamefont {Maunz}, \citenamefont {Olmschenk}, \citenamefont
  {Younge}, \citenamefont {Matsukevich}, \citenamefont {Duan},\ and\
  \citenamefont {Monroe}}]{Moehring2007}%
  \BibitemOpen
  \bibfield  {author} {\bibinfo {author} {\bibfnamefont {D.~L.}\ \bibnamefont
  {Moehring}}, \bibinfo {author} {\bibfnamefont {P.}~\bibnamefont {Maunz}},
  \bibinfo {author} {\bibfnamefont {S.}~\bibnamefont {Olmschenk}}, \bibinfo
  {author} {\bibfnamefont {K.~C.}\ \bibnamefont {Younge}}, \bibinfo {author}
  {\bibfnamefont {D.~N.}\ \bibnamefont {Matsukevich}}, \bibinfo {author}
  {\bibfnamefont {L.~M.}\ \bibnamefont {Duan}}, \ and\ \bibinfo {author}
  {\bibfnamefont {C.}~\bibnamefont {Monroe}},\ }\href
  {http://dx.doi.org/10.1038/nature06118} {\bibfield  {journal} {\bibinfo
  {journal} {Nature}\ }\textbf {\bibinfo {volume} {449}},\ \bibinfo {pages}
  {68} (\bibinfo {year} {2007})}\BibitemShut {NoStop}%
\bibitem [{\citenamefont {Li}\ \emph {et~al.}(2008)\citenamefont {Li},
  \citenamefont {Chalapat},\ and\ \citenamefont {Paraoanu}}]{Li2008}%
  \BibitemOpen
  \bibfield  {author} {\bibinfo {author} {\bibfnamefont {J.}~\bibnamefont
  {Li}}, \bibinfo {author} {\bibfnamefont {K.}~\bibnamefont {Chalapat}}, \ and\
  \bibinfo {author} {\bibfnamefont {G.~S.}\ \bibnamefont {Paraoanu}},\ }\href
  {\doibase 10.1103/PhysRevB.78.064503} {\bibfield  {journal} {\bibinfo
  {journal} {Phys. Rev. B}\ }\textbf {\bibinfo {volume} {78}},\ \bibinfo
  {pages} {064503} (\bibinfo {year} {2008})}\BibitemShut {NoStop}%
\bibitem [{\citenamefont {Julsgaard}\ and\ \citenamefont
  {M\o{}lmer}(2012)}]{Julsgaard2012}%
  \BibitemOpen
  \bibfield  {author} {\bibinfo {author} {\bibfnamefont {B.}~\bibnamefont
  {Julsgaard}}\ and\ \bibinfo {author} {\bibfnamefont {K.}~\bibnamefont
  {M\o{}lmer}},\ }\href {\doibase 10.1103/PhysRevA.85.032327} {\bibfield
  {journal} {\bibinfo  {journal} {Phys. Rev. A}\ }\textbf {\bibinfo {volume}
  {85}},\ \bibinfo {pages} {032327} (\bibinfo {year} {2012})}\BibitemShut
  {NoStop}%
\bibitem [{\citenamefont {Roch}\ \emph {et~al.}(2014)\citenamefont {Roch},
  \citenamefont {Schwartz}, \citenamefont {Motzoi}, \citenamefont {Macklin},
  \citenamefont {Vijay}, \citenamefont {Eddins}, \citenamefont {Korotkov},
  \citenamefont {Whaley}, \citenamefont {Sarovar},\ and\ \citenamefont
  {Siddiqi}}]{Roch2014}%
  \BibitemOpen
  \bibfield  {author} {\bibinfo {author} {\bibfnamefont {N.}~\bibnamefont
  {Roch}}, \bibinfo {author} {\bibfnamefont {M.~E.}\ \bibnamefont {Schwartz}},
  \bibinfo {author} {\bibfnamefont {F.}~\bibnamefont {Motzoi}}, \bibinfo
  {author} {\bibfnamefont {C.}~\bibnamefont {Macklin}}, \bibinfo {author}
  {\bibfnamefont {R.}~\bibnamefont {Vijay}}, \bibinfo {author} {\bibfnamefont
  {A.~W.}\ \bibnamefont {Eddins}}, \bibinfo {author} {\bibfnamefont {A.~N.}\
  \bibnamefont {Korotkov}}, \bibinfo {author} {\bibfnamefont {K.~B.}\
  \bibnamefont {Whaley}}, \bibinfo {author} {\bibfnamefont {M.}~\bibnamefont
  {Sarovar}}, \ and\ \bibinfo {author} {\bibfnamefont {I.}~\bibnamefont
  {Siddiqi}},\ }\href {\doibase 10.1103/PhysRevLett.112.170501} {\bibfield
  {journal} {\bibinfo  {journal} {Phys. Rev. Lett.}\ }\textbf {\bibinfo
  {volume} {112}},\ \bibinfo {pages} {170501} (\bibinfo {year}
  {2014})}\BibitemShut {NoStop}%
\bibitem [{\citenamefont {Sitek}\ and\ \citenamefont
  {Manolescu}(2013)}]{Sitek2013}%
  \BibitemOpen
  \bibfield  {author} {\bibinfo {author} {\bibfnamefont {A.}~\bibnamefont
  {Sitek}}\ and\ \bibinfo {author} {\bibfnamefont {A.}~\bibnamefont
  {Manolescu}},\ }\href {\doibase 10.1103/PhysRevA.88.043807} {\bibfield
  {journal} {\bibinfo  {journal} {Phys. Rev. A}\ }\textbf {\bibinfo {volume}
  {88}},\ \bibinfo {pages} {043807} (\bibinfo {year} {2013})}\BibitemShut
  {NoStop}%
\bibitem [{\citenamefont {Agarwal}(2013)}]{agarwal2013quantum}%
  \BibitemOpen
  \bibfield  {author} {\bibinfo {author} {\bibfnamefont {G.}~\bibnamefont
  {Agarwal}},\ }\href {https://books.google.co.in/books?id=7KKw\_XIYaioC}
  {\emph {\bibinfo {title} {Quantum Optics}}},\ Quantum Optics\ (\bibinfo
  {publisher} {Cambridge University Press},\ \bibinfo {year}
  {2013})\BibitemShut {NoStop}%
\bibitem [{\citenamefont {Sitek}\ and\ \citenamefont
  {Machnikowski}(2012)}]{Sitek2012}%
  \BibitemOpen
  \bibfield  {author} {\bibinfo {author} {\bibfnamefont {A.}~\bibnamefont
  {Sitek}}\ and\ \bibinfo {author} {\bibfnamefont {P.}~\bibnamefont
  {Machnikowski}},\ }\href {\doibase 10.1103/PhysRevB.86.205315} {\bibfield
  {journal} {\bibinfo  {journal} {Phys. Rev. B}\ }\textbf {\bibinfo {volume}
  {86}},\ \bibinfo {pages} {205315} (\bibinfo {year} {2012})}\BibitemShut
  {NoStop}%
\end{thebibliography}%
\end{document}